\documentclass[3p,times,procedia]{elsarticle}

\usepackage{ecrc}

\volume{00}

\firstpage{1}

\journalname{Procedia Computer Science}

\runauth{Z. Xu et al.}

\jid{procs}

\jnltitlelogo{Procedia Computer Science}

\CopyrightLine{2023}{The Authors. Published by Elsevier B.V.\newline Selection and/or peer-review under responsibility of ITQM 2023}

\usepackage{amssymb}

\usepackage[figuresright]{rotating}

\usepackage{amsfonts}
\usepackage{graphicx,amssymb,mathrsfs,amsmath,color,fancyhdr}
\usepackage{subcaption}
\usepackage{amsthm}
\usepackage{manfnt}
\usepackage{cases}
\usepackage{mathbbol}
\usepackage{setspace}
\newtheorem{theorem}{Theorem}

\newtheorem{corollary}{Corollary}
\newtheorem{proposition}{Proposition}

\begin{document}

\begin{frontmatter}



\dochead{Information Technology and Quantitative Management (ITQM 2023)}

\title{Are Large Traders Harmed by Front-running HFTs?}


\author[a]{Ziyi Xu}
\author[a,1]{Xue Cheng}

\address[a]{LMEQF, Department of Financial Mathematics, School of Mathematical Sciences,
Peking University, Beijing 100871, China.}

\begin{abstract}
This paper studies the influences of a high-frequency trader (HFT) on a large trader whose future trading is predicted by the former.
We conclude that HFT always front-runs and the large trader is benefited when: (1) there is sufficient high-speed noise trading; (2) HFT's prediction is vague enough. Besides, we find surprisingly that (1) making HFT's prediction less accurate might decrease large trader's profit; (2) when there is little high-speed noise trading, although HFT nearly does nothing, the large trader is still hurt.
\end{abstract}


\end{frontmatter}

\correspondingauthor[1]{chengxue@pku.edu.cn}


\section{Introduction}
\label{intro}
Front-runner predicts large traders' incoming orders, trades in the same direction ahead of them, and then closes the positions when they arrive, hoping to profit from the price impacts caused by large traders. For example, if a front-runner asserts that an investor is going to buy, she will buy in advance and then sell back to the investor.

 As found in Kirilenko et al. (2017) \cite{2017The}, in today's market, most front-runners are high-frequency traders (HFTs), who process information faster and send orders with lower latency.  The development of high-frequency trading makes front-running a major concern. 

Conventional wisdom has it that front-runners increase the large trader's costs by taking away cheaper liquidity. However, some studies do believe that under certain conditions, front-runners could be beneficial. Empirical work supports both views, a unified conclusion has not been reached. 

In this paper, through an extended celebrated Kyle's \cite{kyle1985continuous} model, we focus on the influences of an HFT on an informed large trader whose future order is detected by the former.
We prove that it is optimal for HFT to front-run and we outline situations where HFT harms or benefits the large trader. Specifically, the front-running HFT is favorable to the large trader when (1) there is sufficient high-speed noise trading; (2) the high-speed noise trading is inadequate but her prediction is inaccurate enough.

We also explore the influences of noise trading and prediction accuracy on investors' behavior and profits. We prove that HFT trades more aggressively with more high-speed noise trading and a more accurate prediction. For the large trader, her trading intensity and profit also increase with the size of high-speed noise trading. Surprisingly, we find that, when HFT's prediction is vague enough, the large trader can be guaranteed to be unharmed, but her profit tends to decrease as HFT's signal gets noisier, which seems counterintuitive with the common sense that the large trader gets better if her trading intentions are deeper hidden. 

The limit results of the above influences are also investigated. We have drawn two interesting conclusions: (1) when almost all noise traders are slow, HFT nearly does nothing, but the large trader is still harmed; (2) when the prediction is perfect and the size of high-speed noise trading tends to infinity, HFT reaches the max front-running volume, which is only half of the large trader's incoming order.

Besides the case where HFT predicts the large trader's order, we also study the case where HFT can predict the aggregate order flow, i.e., the large trader's order and its accompanying noise orders. From a practical point of view, this kind of HFTs has more advanced technology to estimate the trend of the whole market, rather than only the major market trend. In this case, we find that HFT still front-runs and always benefits the large trader. 

\subsection{Related literature}
Front-running strategy exploits other investors' need to trade, thus belonging to the topic of predatory trading. In the pioneering work Brunnermeier and Pedersen (2005) \cite{brunnermeier2005predatory}, only permanent price impact is considered and front-runners bring inferior prices to the large trader. 
Bessembinder et al. (2016) \cite{bessembinder2016liquidity} considers transient price impact. The authors find that when the market is quite resilient, front-runners could benefit the large trader. A resilient market can be regarded as active and liquid, which is consistent with our discovery that HFT may be good for the large trader in a market with abundant noise trading. Other related works include Carlin et al. (2007) \cite{carlin2007episodic} and Sch{\"o}neborn and Schied (2009) \cite{schoneborn2009liquidation}. In the above works, predatory traders are perfectly aware of large trader's intention, while in this paper, the impacts of prediction accuracy are further investigated.

When it comes to high-frequency trading, Li (2018) \cite{li2018high} models front-running HFTs who predict the sum of informed and noise orders and concludes that informed trader trades less aggressively in the presence of HFTs. 
Yang and Zhu (2020) \cite{yang2020back} models back-runners who use information about trading history to predict the informed traders' future path and trade along with them. The main difference between this paper and the above works is that different-size high-speed noise trading is considered, which is proved to be crucial for HFT's influences. For a more thorough review of high-frequency trading literature, readers could refer to Menkveld (2016) \cite{menkveld2016economics}. 

Consequently, we contribute to the front-running literature by studying its influences on the large trader under various market conditions and different prediction accuracy, and some unexpected results are obtained.

\section{The Model and Equilibrium}
We start by introducing the following two-period model ($t=0,1,2$) for front-running, which is a variant of the classic Kyle's model.

In this market, a risky asset is traded whose true value, $v$, is normally distributed as
$$
    v\sim N(p_0,\sigma_v^2).
$$

There are four types of market participants: (1) \textit{dealers}, who are assumed to be competitive and risk-neutral; (2) a normal-speed large \textit{informed trader} (IT, for short), who privately knows $v$; (3) a \textit{High-Frequency Trader} (HFT, for short), who can get a signal about IT's future trading and might act as a front-runner; (4) \textit{noise traders}, who trade randomly. 

 At $t=0,$ IT submits a market order of quantity $i=i(v)$, based on her private knowledge. However, for some reasons, e.g., the delay in sending orders, as mentioned in \cite{li2018high}, \cite{baldauf2020high}, the order is not executed until $t=2$.

Very soon after IT sends the order, HFT gets a signal about $i$:
$$
    \hat{i}=i+z,
$$
where the signal noise $z$ is independent of $i$ and follows $ N(0,\sigma_z^2)$. 

At $t=1$, HFT sends a market order of quantity $x=x(\hat{i})$. We assume that HFT's orders are always fulfilled at once. HFT will offset this position at $t=2$ by sending a market order of size $-x.$

The aggregate noise orders in each period are denoted by $u_1$ and $u_2$, where
$$u_1\sim N(0,\sigma_1^2),\ u_2\sim N(0,\sigma_2^2),$$ are independent of each other and any other random variables. 

To sum up, the total order flow $y_1$ and $y_2$ at $t=1$ and $t=2$ respectively are
$$
    y_1=x+u_1,\ y_2=i+u_2-x.
$$

 In the following, we assume $\sigma_2>0$ as in Kyle's model and use
$$
    \theta_z=\frac{\sigma_z^2}{\sigma_2^2},\ \theta_1=\frac{\sigma_1^2}{\sigma_2^2}
$$
to characterize the relative signal accuracy and the relative size of high-speed noise trading.

In equilibrium, competitive and risk-neutral dealers set prices according to the weak-efficiency rule:
    $$
\begin{aligned}
p_1=\mathbb{E}(v|y_1),\ p_2=\mathbb{E}(v|y_1,y_2).
\end{aligned}
    $$
Both IT and HFT seek to maximize their expected profit:
\begin{equation*}
i^*=\arg\max_{i=i(v)}\mathbb{E}\big((v-p_2)i\big|v\big),
    \end{equation*}
\begin{equation*}
x^*=\arg\max_{x=x(\hat{i})}\mathbb{E}\big((p_2-p_1)x\big|\hat{i}\big).
    \end{equation*}


\section{Main Results}
Since there is no general equilibrium for $\theta_1>0$ and $\theta_1=0$, we discuss the two cases separately. We assume $p_0=0$, for it does not bring out any essential changes in results. 
\subsection{Equilibrium with high-speed noise trading}
\label{sectionmain}
\begin{theorem}
\label{mainthm}
Given $\theta_1>0, \theta_z\geq0$, there exists a unique equilibrium $\{p_1,p_2,i^*,x^*\}$, 
where HFT follows the strategy $x^*=\beta^*\hat{i}$ and $\beta^*\in(0,1)$ solves the equation
\begin{equation}
\label{betastar}
\begin{aligned}
 0=& \beta^6(4\theta_1\theta_z^2+\theta_1\theta_z^3+2\theta_1^2\theta_z^2+2\theta_z^2+\theta_z^3)+\beta^5(4\theta_1\theta_z+4\theta_1\theta_z^2+2\theta_1\theta_z^3+8\theta_1^2\theta_z+4\theta_1^2\theta_z^2+4\theta_1^3\theta_z)\\
+ & \beta^4(2\theta_1\theta_z+\theta_1\theta_z^2-11\theta_1^2\theta_z-8\theta_1^2\theta_z^2-13\theta_1^3\theta_z) 
+\beta^3(2\theta_1^2 + 2\theta_1^3 + 8\theta_1^2\theta_z + 4\theta_1^2\theta_z^2+ 16\theta_1^3\theta_z ) \\
  - & \beta^2( \theta_1^2\theta_z +5\theta_1^3 + 9\theta_1^3\theta_z) + \beta(4\theta_1^3 + 2\theta_1^3\theta_z) -\theta_1^3;
\end{aligned}
\end{equation}
IT follows the strategy $i^*=\alpha^* v$ and
\begin{equation*}
\label{alphastar}
    \alpha^*=\frac{\sigma_2}{\sigma_v}\sqrt{\frac{\theta_1+\beta^{*2}\theta_z(\theta_1+1)}{\theta_1(1-\beta^*)^2+\beta^{*2}(\theta_z+1)}};
\end{equation*}
the liquidation price at time 1 is
\begin{equation*}
\begin{aligned}
&p_1=\lambda_1^* (x^*+u_1),\\
&\lambda_1^*=\frac{\sigma_v}{2\sigma_2}\frac{2\beta^*\sqrt{(\beta^{*2}\theta_z(\theta_1+1)+\theta_1)(\theta_1(1-\beta^*)^2+\beta^{*2}(\theta_z+1))}}{\beta^{*2}(\beta^{*2}\theta_z(\theta_1+1)+\theta_1)+(\beta^{*2}\theta_z+\theta_1)(\theta_1(1-\beta^*)^2+\beta^{*2}(\theta_z+1))};
\end{aligned}
\end{equation*}
the liquidation price at time 2 is
\begin{equation*}
\begin{aligned}
&p_2=\mu_1^*(x^*+u_1)+\mu_2^*(i^*+u_2-x^*),\\
&\mu_1^*=\frac{\sigma_v}{2\sigma_2}\frac{\beta^{*2}\theta_z+\beta^*}{\sqrt{(\theta_1(1-\beta^*)^2+\beta^{*2}(\theta_z+1))(\theta_1+\beta^{*2}\theta_z(\theta_1+1))}},\\
&\mu_2^*=\frac{\sigma_v}{2\sigma_2}\frac{\beta^{*2}\theta_z+(1-\beta^*)\theta_1}{\sqrt{(\theta_1(1-\beta^*)^2+\beta^{*2}(\theta_z+1))(\theta_1+\beta^{*2}\theta_z(\theta_1+1))}}.
\end{aligned}
\end{equation*}
\end{theorem}

From Theorem \ref{mainthm}, we see that in equilibrium $\beta^*\in(0,1),$ which means that HFT trades in the same direction as IT at $t=1$ and against her at $t=2,$ i.e., HFT front-runs. With HFT, $\mathbb{E}(\pi^{\text{IT}})=\frac{\sigma_v^2}{2}\alpha^*$; 
without HFT, $\mathbb{E}(\pi^{\text{IT}}_0)=\frac{\sigma_v\sigma_2}{2}.$ So IT is benefited by HFT if $\alpha^*>\frac{\sigma_2}{\sigma_v}.$ We naturally wonder when it happens.
\begin{theorem}
\label{benefit}
IT is favored by HFT when $(\theta_1,\theta_z)$ is in the following areas:
\begin{equation*}
\begin{aligned}
&\theta_1>\frac{2\sqrt{3}-3}{3},\ \theta_z\geq0;    \\
&0<\theta_1\leq\frac{2\sqrt{3}-3}{3},\ \theta_z>\overline{\theta}_z=\frac{-(\theta_1+5)+2\sqrt{4\theta_1^2+10\theta_1+5}}{-5\theta_1}.
\end{aligned}
\end{equation*}
 Otherwise, IT is harmed.
\end{theorem}
When $(\theta_1,\theta_z)$ is in the first area, HFT cannot harm IT even if her signal is perfectly accurate since the market provides enough noise shelter for IT. When $(\theta_1,\theta_z)$ is in the second area, IT is benefited when HFT's signal is relatively vague. What's more,
larger the $\theta_1$, smaller the $\overline{\theta}_z$. That is, as the size of noise trading gets larger, less signal noise is needed to disturb HFT.

In equilibrium, the optimal intensities of investors, $\alpha^*$ and $\beta^*$, both depend on  $\theta_1$ and $\theta_z$. Now we investigate how $\alpha^*$ and $\beta^*$ change with them.

\begin{proposition}
\label{HFTaction}
The optimal intensity of HFT, $\beta^*$, increases with $\theta_1$ and decreases with $\theta_z$.
\end{proposition}

A larger $\theta_1$ means more noise orders accompanying HFT's fast trading, which enables HFT to employ more of her speed priority while bear less impact. A larger $\theta_z$ means a less accurate signal, which makes HFT less confident about IT's future trading, and hence trades more conservatively. 

The impacts of $\theta$s on IT's behavior are more complicated:
\begin{proposition}
\label{ITaction}
\begin{description}
\item[(1).] $\alpha^*$ increases with $\theta_1$;
    \item[(2).] 
given $\theta_1\geq\frac{1}{2},\alpha^*$ decreases with $\theta_z$;
\item[(3).] given $0<\theta_1<\frac{1}{2},$ $\alpha^*$ increases with $\theta_z$ when $0\leq\theta_z<\Tilde{\theta}_z$  and decreases with it when $\theta_z\geq\Tilde{\theta}_z$, where
\begin{equation*}
    \Tilde{\theta}_z=\frac{1-\theta_1-2\theta_1^2}{3\theta_1}.
\end{equation*}
\end{description}
\end{proposition}

We surprisingly find that when $\theta_1\geq\frac{1}{2}$, $\alpha^*$, and consequently, IT's expected profit, decrease with the size of signal noise. It actually answers a controversial question:
is HFT's having more precise information detrimental to other investors? From our observation, it depends on the size of market noise, if high-speed noise trading is active enough, it is not necessary to increase the noise in HFT's signal.

Since the equilibrium has good analytical properties, we can investigate the limit results of the above influences.

\begin{proposition}
\label{theta1limit}
Given $\theta_z\geq0$, when $\theta_1\rightarrow\infty,$ $\beta^*$ converges to the root of the following equation:
\begin{equation*}
4\theta_z\beta^5-13\theta_z\beta^4+(2+16\theta_z)\beta^3-(5+9\theta_z)\beta^2+(4+2\theta_z)\beta-1=0;
\end{equation*}
the optimal intensity of IT, 
\begin{equation}
\label{alphatheta1infty}
    \alpha^*\rightarrow\frac{\sigma_2}{\sigma_v}\sqrt{\frac{\beta^{*2}\theta_z+1}{(1-\beta^*)^2}}>\frac{\sigma_2}{\sigma_v}.
\end{equation}
When $\theta_1\rightarrow0$, $\beta^*\rightarrow0,$ 
\begin{equation}
\label{alphatheta10}
\begin{aligned}
&\alpha^*\rightarrow\frac{\sigma_2}{\sigma_v}\sqrt{\frac{y^2\theta_z^{2x}+\theta_z}{y^2\theta_z^{2x}+\theta_z+1}}<\frac{\sigma_2}{\sigma_v},\\
\end{aligned}
\end{equation}
where $x=0.3245,y=1.3845.$
\end{proposition}
When $\theta_1$ is large enough, from \eqref{alphatheta1infty},  IT trades and profits more with an HFT. When nearly all noise traders are normal-speed ($\theta_1\rightarrow0$), HFT almost does nothing ($\beta^*\rightarrow0$), however, from \eqref{alphatheta10}, IT is still worse off.

Combing Proposition \ref{HFTaction}, \ref{ITaction} and \ref{theta1limit}, we have the following intersting corollary:
\begin{corollary}
Overall parameters $\theta_1>0,\theta_z\geq0,$
\begin{equation*}
\beta^*(\theta_1,\theta_z)\leq\beta^*(\theta_1=\infty,\theta_z=0)=\frac{1}{2},
\end{equation*}
\begin{equation*}
\alpha^*(\theta_1,\theta_z)\leq\alpha^*(\theta_1=\infty,\theta_z=0)=2\frac{\sigma_2}{\sigma_v}.
\end{equation*}
\end{corollary}
It implies that (1) HFT will not front-run more than half of the informed order; (2) IT's intensity $\alpha^*$ and expected profit are at most twice of those without HFT.

\begin{proposition}
\label{thetazinfty}
Given $\theta_1>0$, when $\theta_z\rightarrow0,$ the equilibrium converges the case $\theta_z=0;$ when $\theta_z\rightarrow\infty$, 
\begin{equation*}
\begin{aligned}
  & \beta^*\rightarrow0,\alpha^*\rightarrow\frac{\sigma_2}{\sigma_v},\mathbb{E}(\pi^{\text{IT}})\rightarrow\frac{\sigma_2\sigma_v}{2},\mathbb{E}(\pi^{\text{HFT}})\rightarrow0,\\ 
   & \lambda_1^*\rightarrow0,\mu_1^*\rightarrow0,\mu_2^*\rightarrow\frac{\sigma_v}{2\sigma_2},\lambda_2^*\rightarrow\frac{\sigma_v}{2\sigma_2}.
\end{aligned}
\end{equation*}
\end{proposition}
From Proposition \ref{thetazinfty}, when HFT predicts nearly nothing, the market converges to the one without her.

\subsection{Equilibrium without high-speed noise trading}
The equilibrium does not exist for $\theta_1=0$. However, we can investigate the partial equilibrium between dealers and IT given HFT's strategy $x=\beta\hat{i}$, which satisfies the weak-efficiency and optimization condition. If HFT does front-run, $i^*=\alpha^* v$ and
\begin{equation*}
\label{alpha*=0}
    \alpha^*=\frac{\sigma_2}{\sigma_v}\sqrt{\frac{\theta_z}{\theta_z+1}}<\frac{\sigma_2}{\sigma_v}.
\end{equation*}
At this time,
HFT's expected profit is zero.
Due to the absence of high-speed noise trading, HFT's front-running causes so large a price impact that harms IT without benefiting herself. When $\beta=0,$ undoubtedly, $\alpha^*=\frac{\sigma_2}{\sigma_v}$, which is in line with the classic result.

\section{Numerical Results}
In this section, we focus on the case $\theta_1>0$. The sixth-order polynomial equation \eqref{betastar} hardly has explicit solutions, so we study investors' behavior through numerical methods. IT's expected profit is omitted since it is linear in $\alpha^*:\mathbb{E}(\pi^{\text{IT}})=\frac{\sigma_v^2}{2}\alpha^*.$ Given $\sigma_v$ and $\sigma_2$, we investigate:
\begin{equation*}
\label{index}
 \beta^*,\ \frac{\mathbb{E}(\pi^{\text{HFT}})}{\sigma_v\sigma_2/2},\ \frac{\alpha^*}{\sigma_2/\sigma_v},
\end{equation*}
which are values only decided by $\theta_1,\theta_z$. In the following figures, we use orange lines to represent the corresponding results in the situation without HFT.

We first do comparative static analyses w.r.t $\theta_1$:
\begin{equation*}
\theta_1\in[10^{-6},10^{-4}],\   \theta_1\in[10^{-4},10^{-2}],\   \theta_1\in[10^{-2},25].
\end{equation*}
Since the general shapes of equilibrium are similar for different $\theta_z,$ we only display results for $\theta_z=0.04$.

\begin{figure}[!htbp]
    \centering
    \subcaptionbox{HFT's action.}{
    \includegraphics[width = 0.27\textwidth]{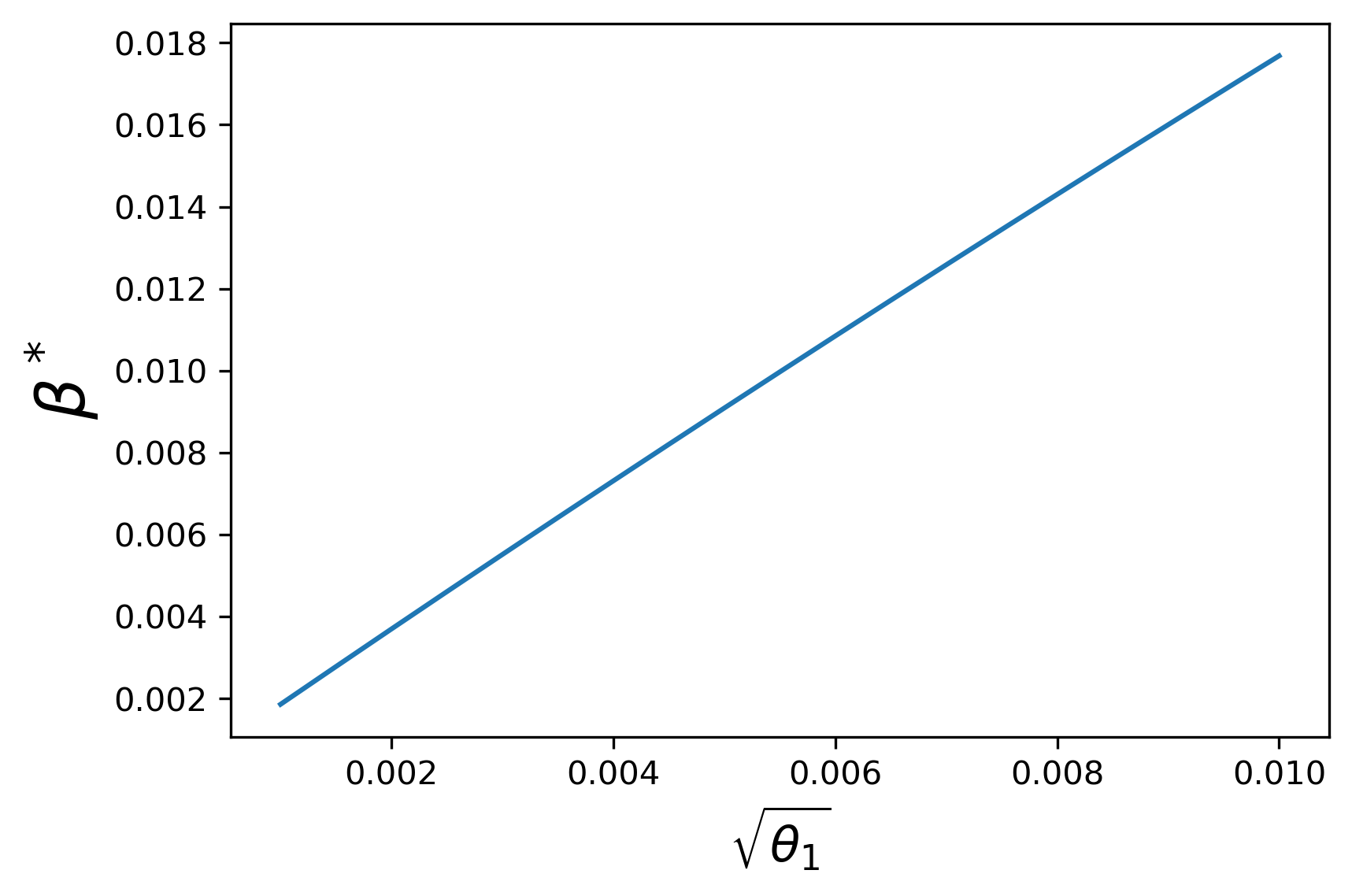}
    }
    \subcaptionbox{HFT's action.}{
    \includegraphics[width = 0.27\textwidth]{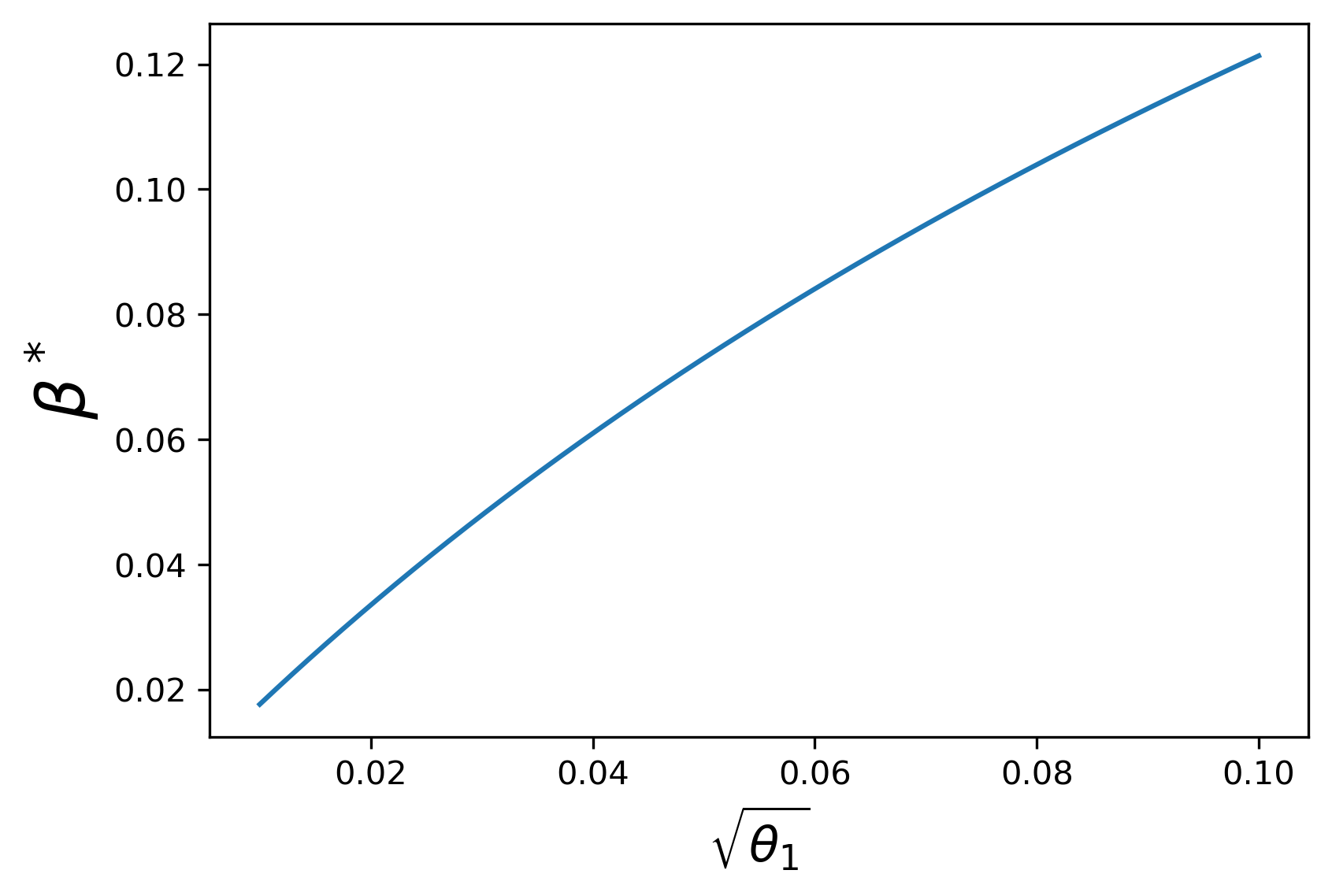}
    }
    \subcaptionbox{HFT's action.}{
    \includegraphics[width = 0.27\textwidth]{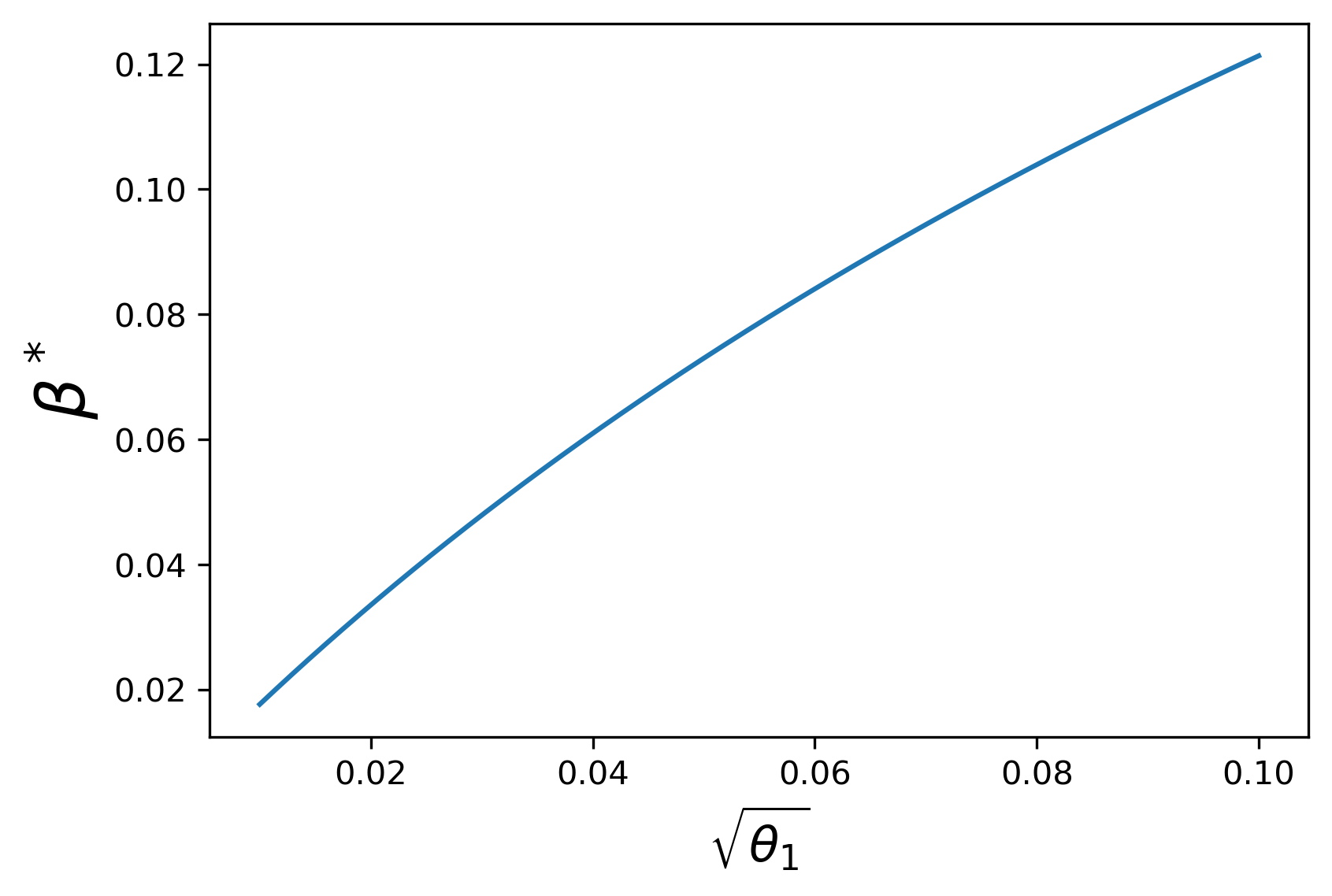}
    }
    
    \subcaptionbox{HFT's profit.}{
    \includegraphics[width = 0.27\textwidth]{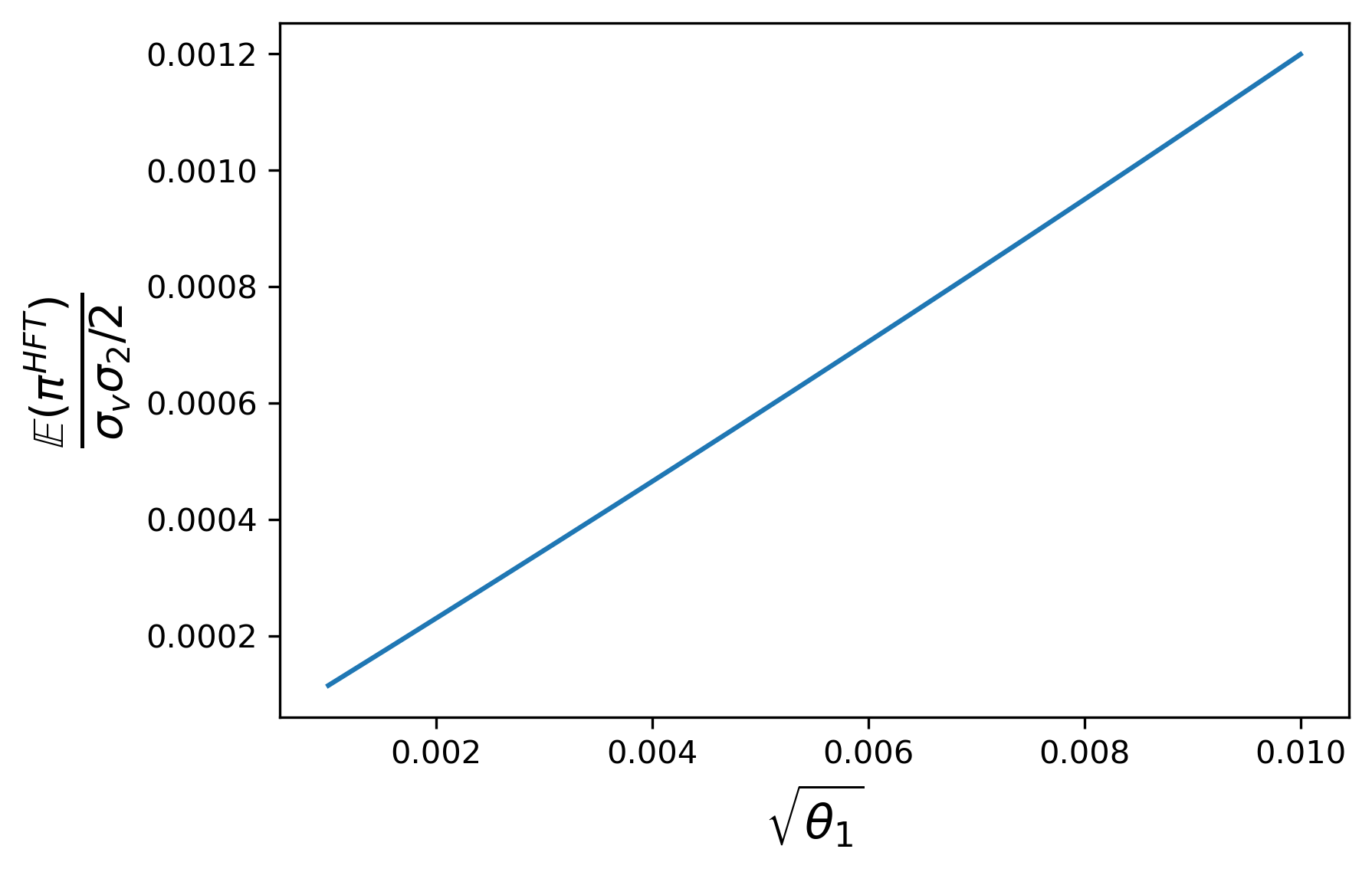}
    }
    \subcaptionbox{HFT's profit.}{
    \includegraphics[width = 0.27\textwidth]{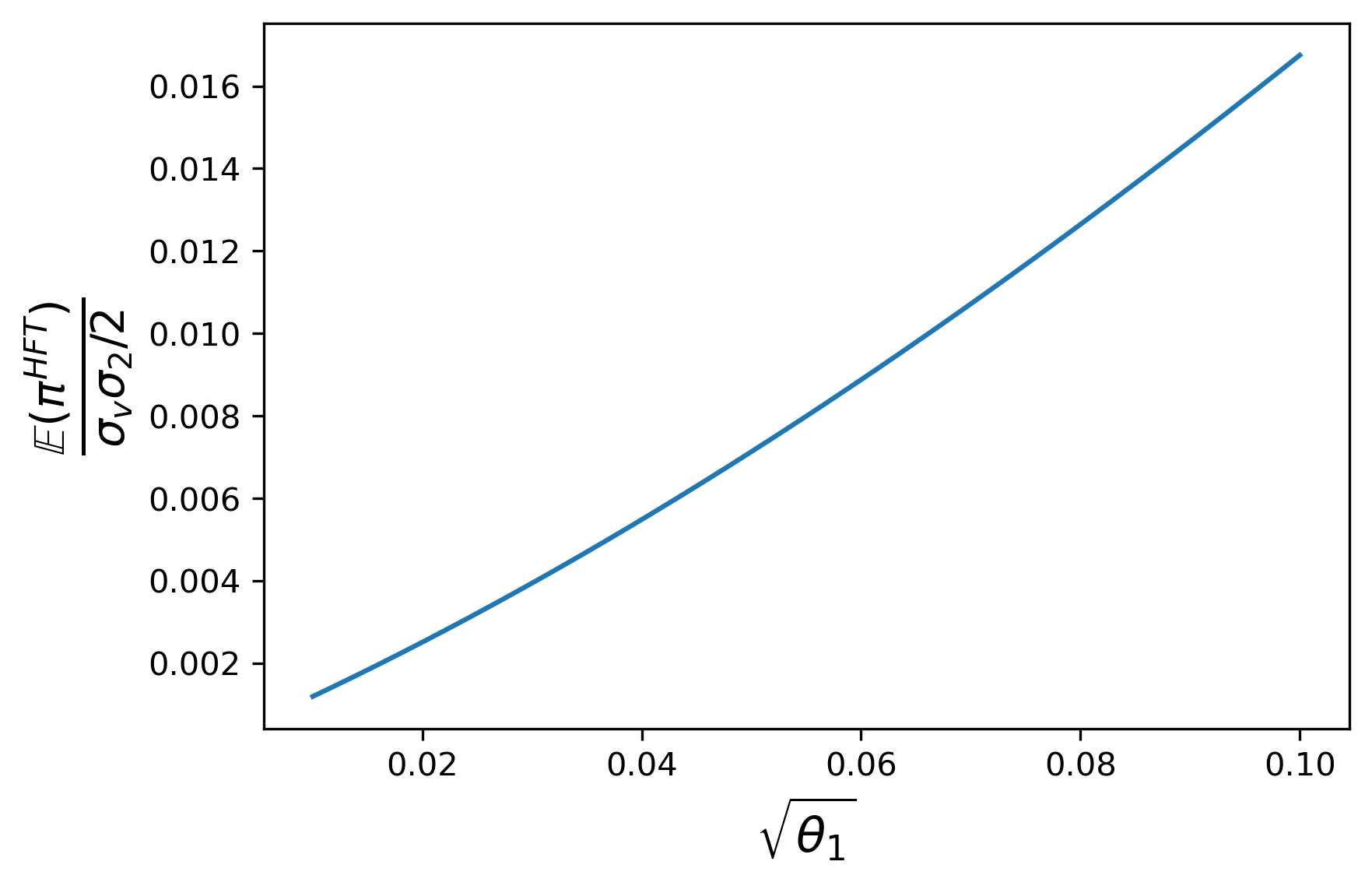}
    }
    \subcaptionbox{HFT's profit.}{
    \includegraphics[width = 0.27\textwidth]{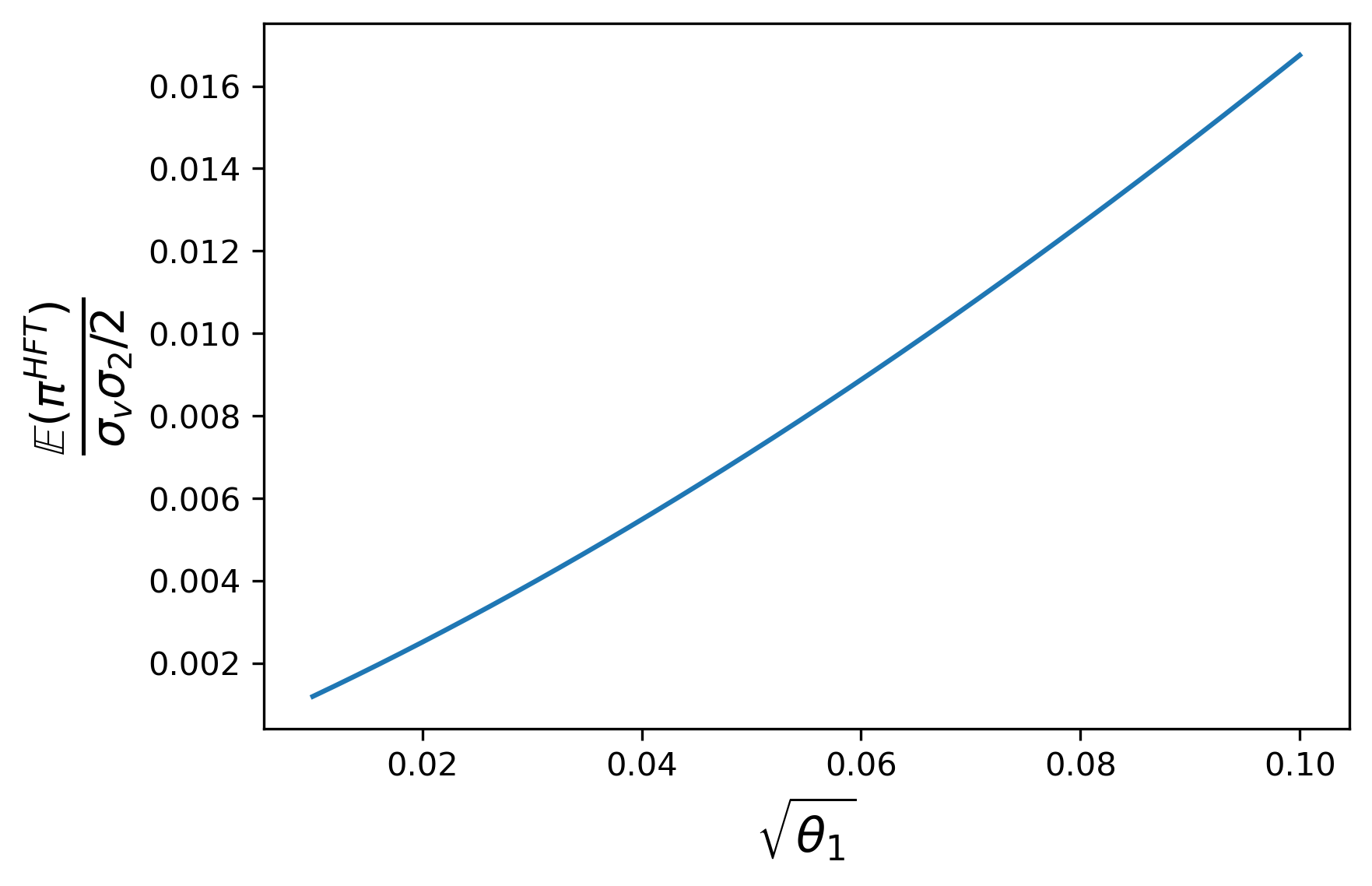}
    }
    \caption{How HFTs' actions and profits change with $\theta_1$.}
    \label{figtheta1-HFT}
\end{figure}

For HFT, $\beta^*$ and her profit increase with $\theta_1$, as shown in Figure \ref{figtheta1-HFT}. HFT is favored by more noise traders since $u_1$ plays the role of a shield, enabling her to employ the prediction more intensely.

\begin{figure}[!htbp]
    \centering
\subcaptionbox{IT's action.}{
    \includegraphics[width = 0.27\textwidth]{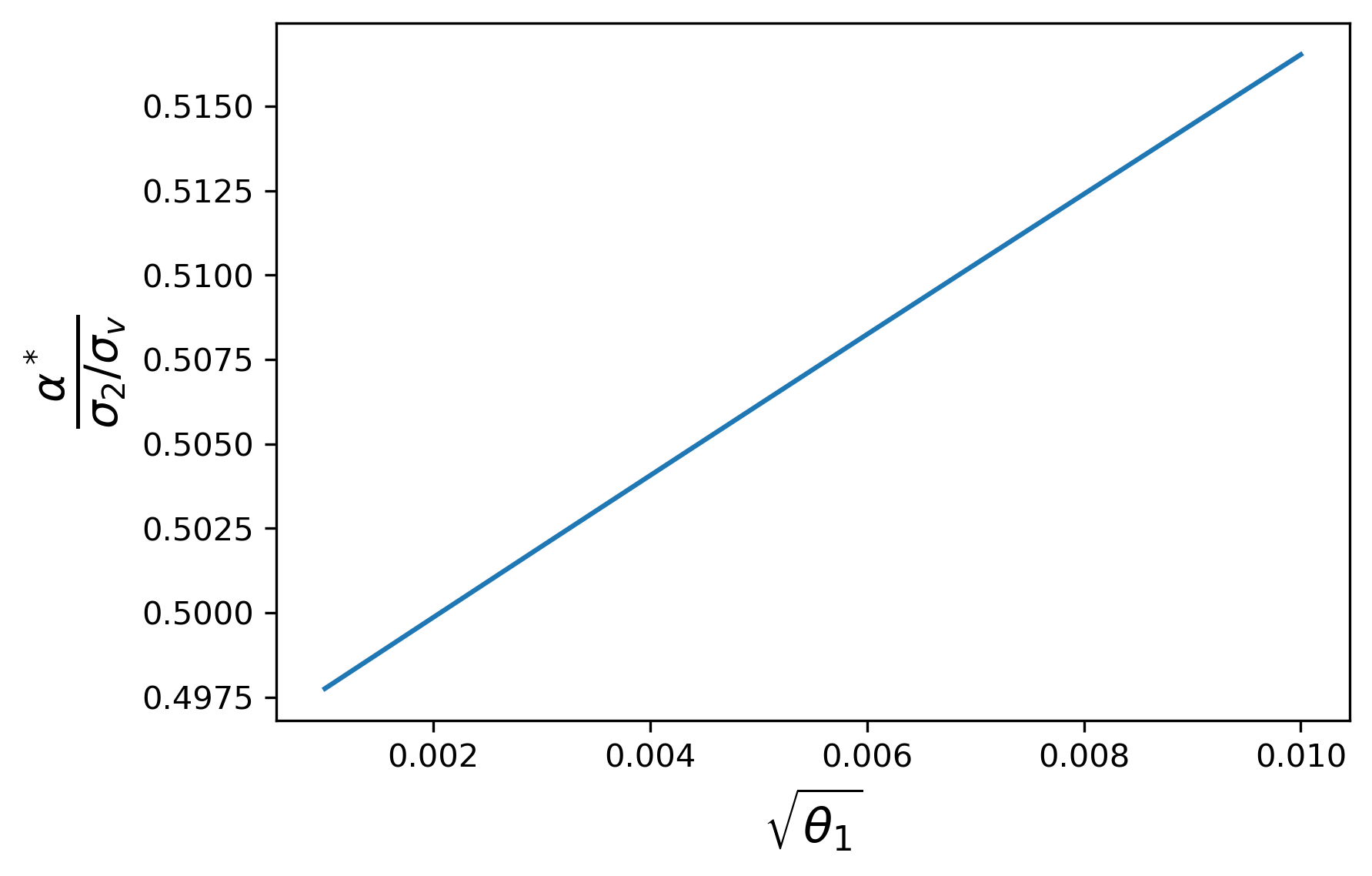}
    }
    \subcaptionbox{IT's action.}{
    \includegraphics[width = 0.27\textwidth]{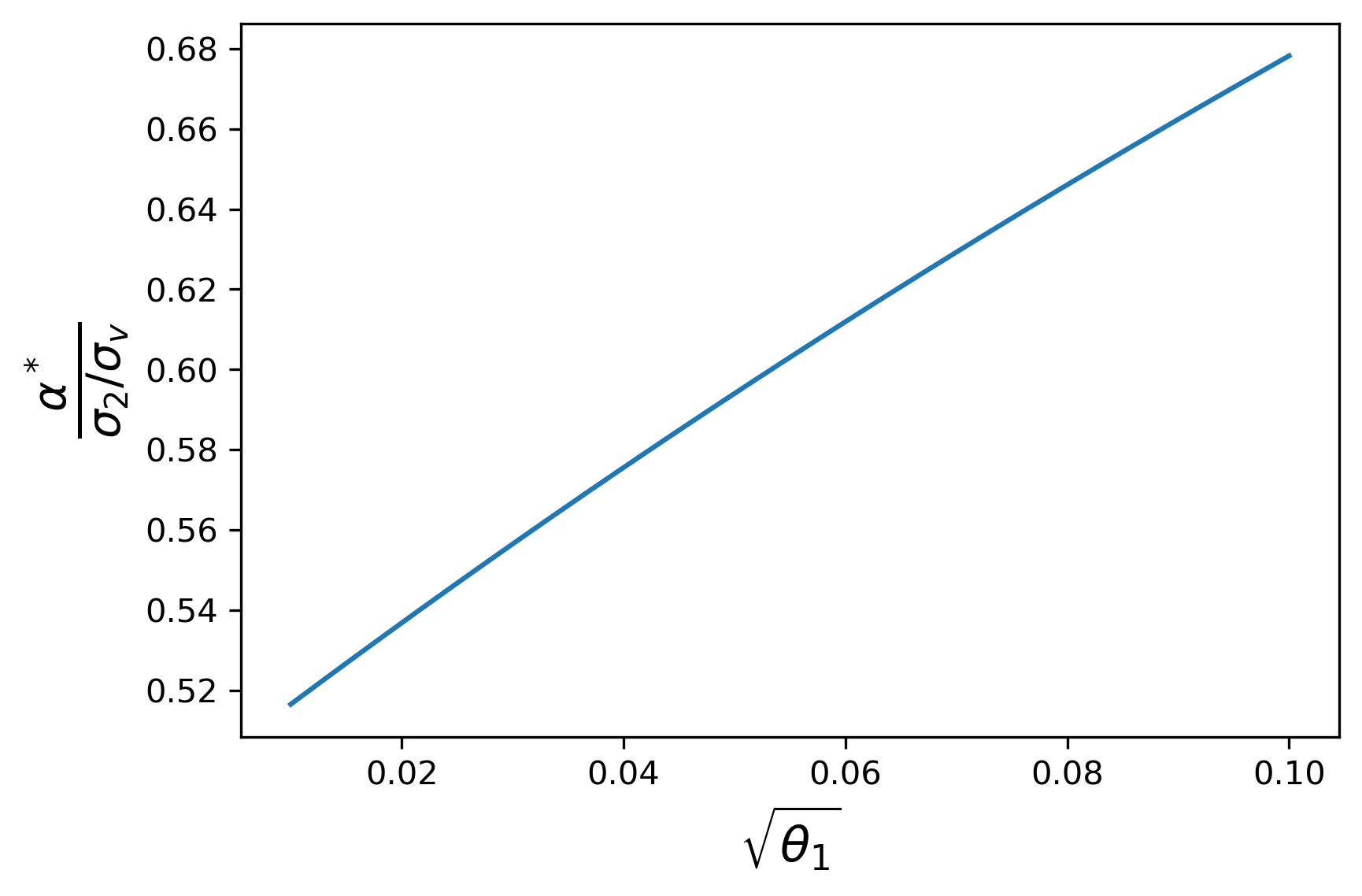}
    }
    \subcaptionbox{IT's action.}{
    \includegraphics[width = 0.27\textwidth]{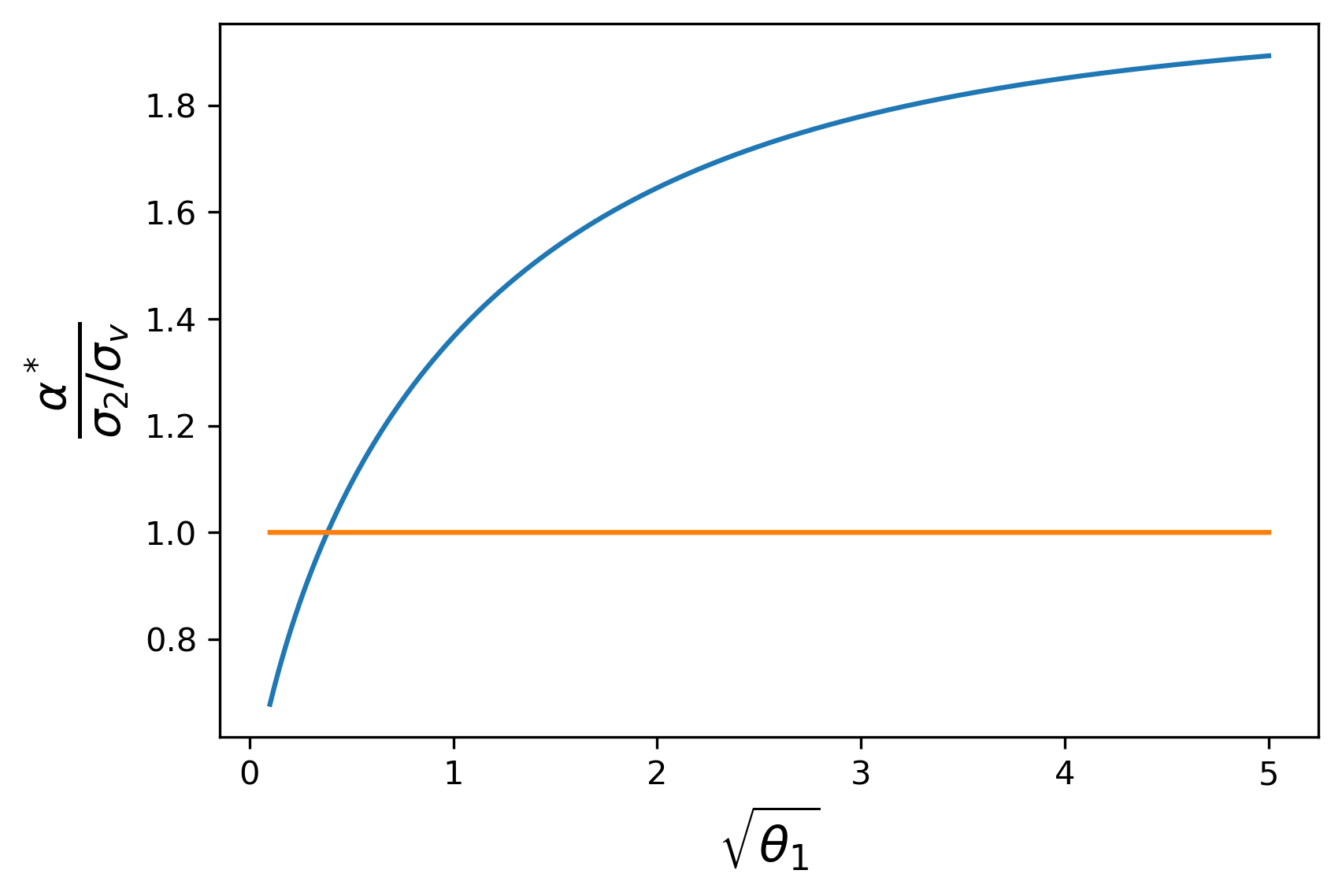}
    }
    \caption{How IT's actions (profits) change with $\theta_1$.}
    \label{figtheta1-IT}
\end{figure}

For IT, she also profits more as $\theta_1$ increases, as shown in Figure \ref{figtheta1-IT}. On the one hand, a larger $\theta_1$ brings a larger $\beta^*$, which increases time-1 price impact, but this impact is also decreased by more noise trading. On the other hand, a larger $\beta^*$ means that HFT provides more liquidity back at time 2. On average, IT trades more aggressively with the greater size of high-speed noise trading.
When $\theta_1$ is large enough ($\theta_1\geq0.15$), the advantages of a large $\theta_1$ outweigh the disadvantages, thus IT is benefited by HFT.

\begin{figure}[!htbp]
    \centering
    \subcaptionbox{HFT's action, $\theta_1=0.12$.}{
    \includegraphics[width = 0.27\textwidth]{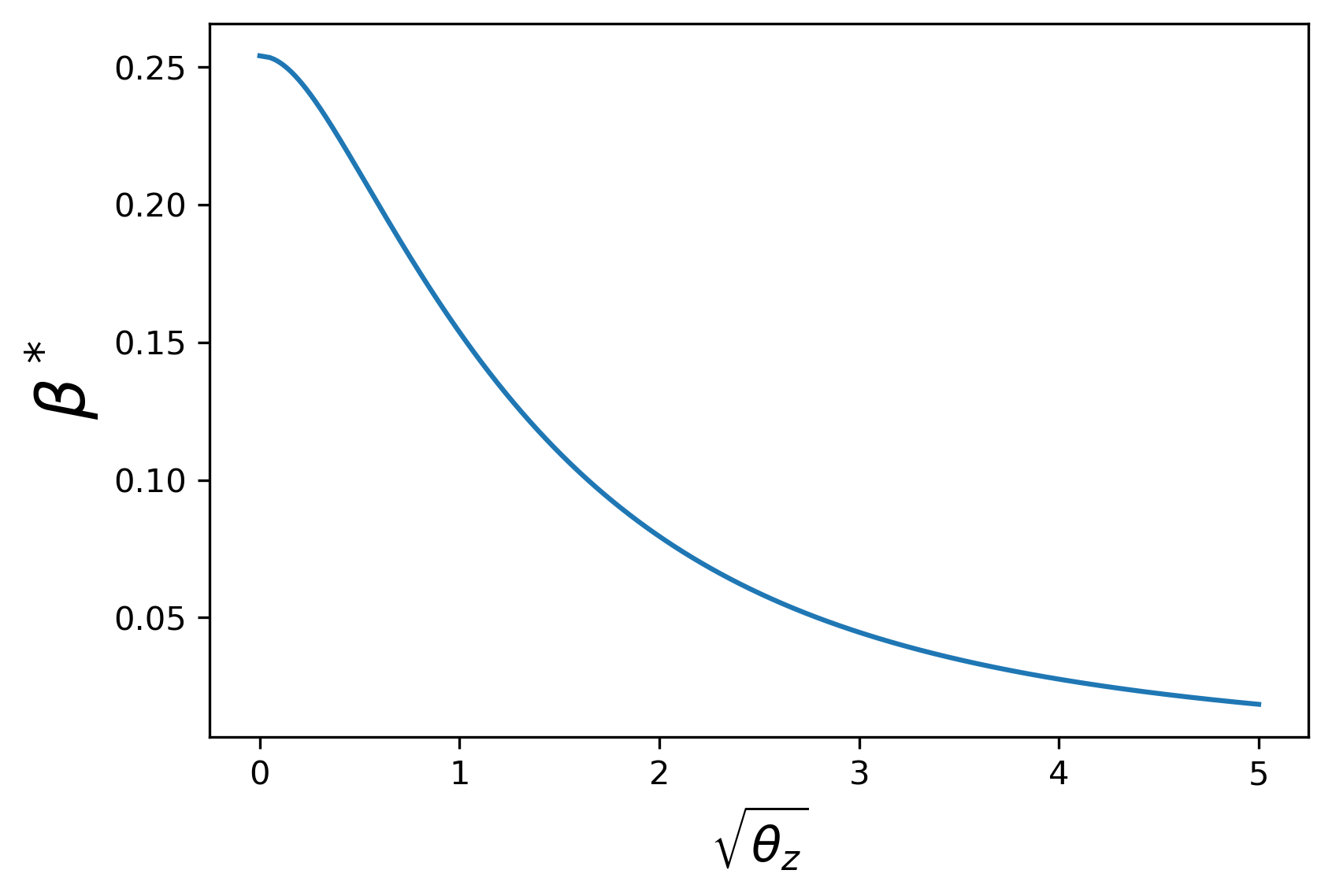}
    }
   \subcaptionbox{HFT's action, $\theta_1=0.2$.}{
    \includegraphics[width = 0.27\textwidth]{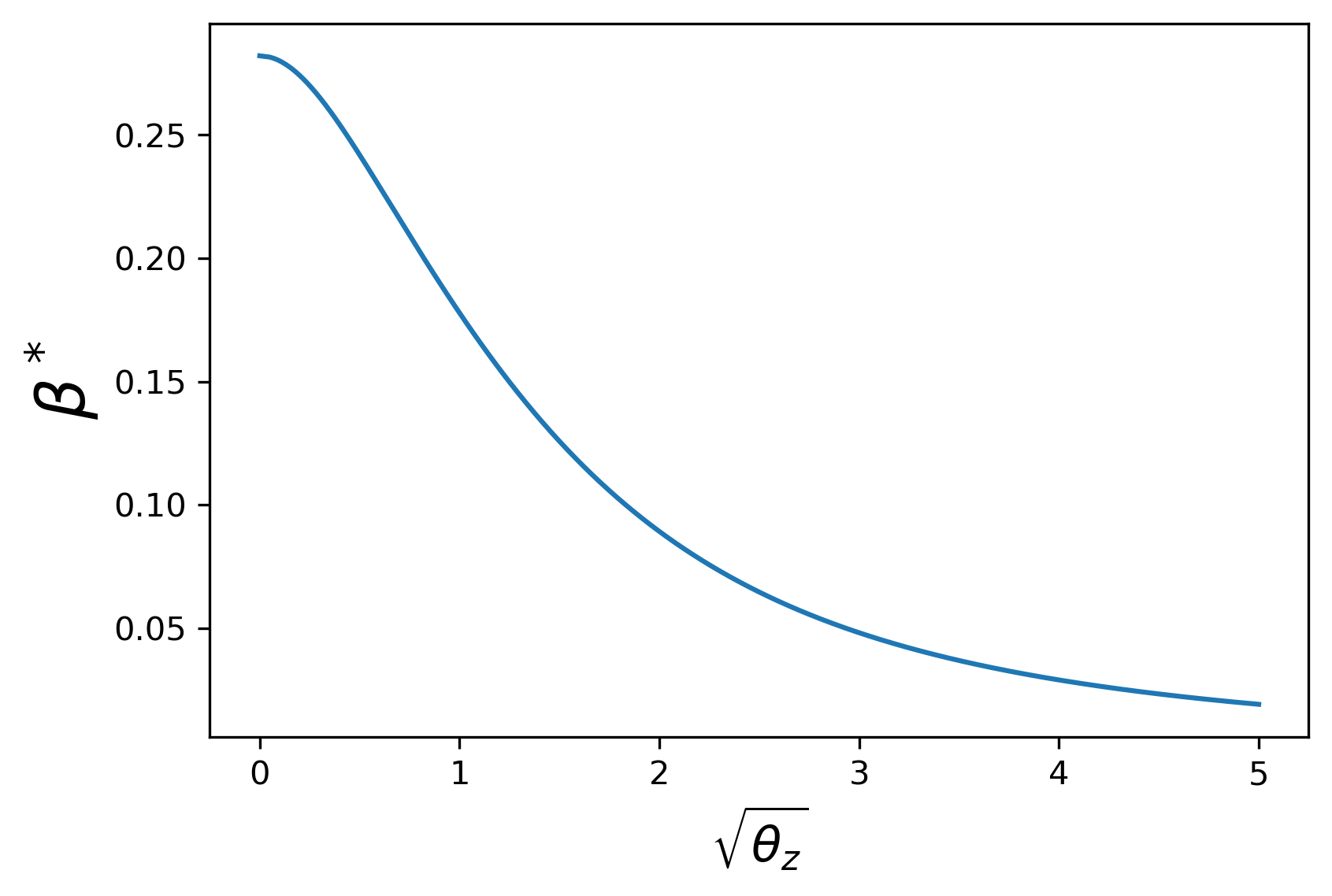}
    }
   \subcaptionbox{HFT's action, $\theta_1=1$.}{
    \includegraphics[width = 0.27\textwidth]{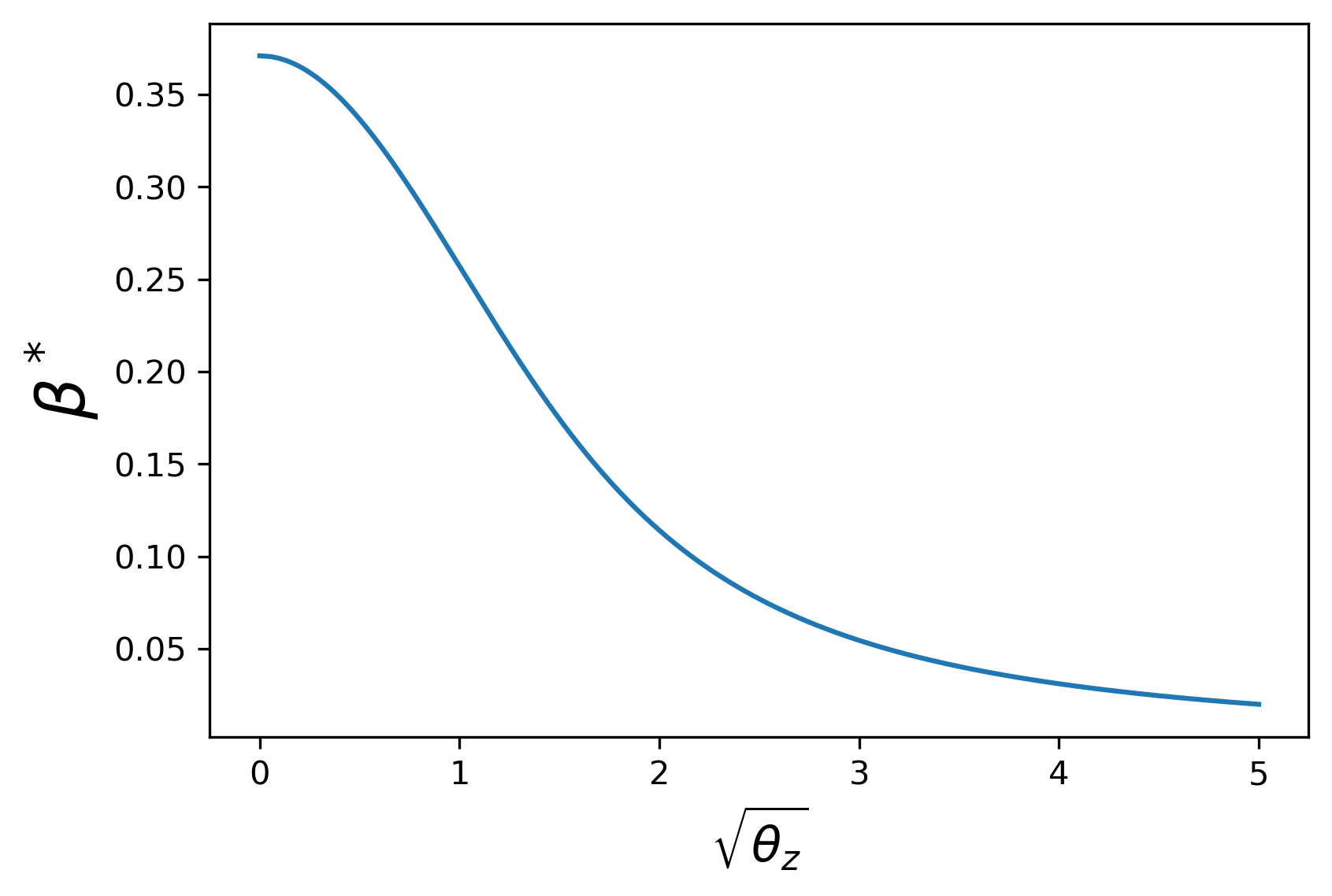}
    }
    
    \subcaptionbox{HFT's profit, $\theta_1=0.12$.}{
    \includegraphics[width = 0.27\textwidth]{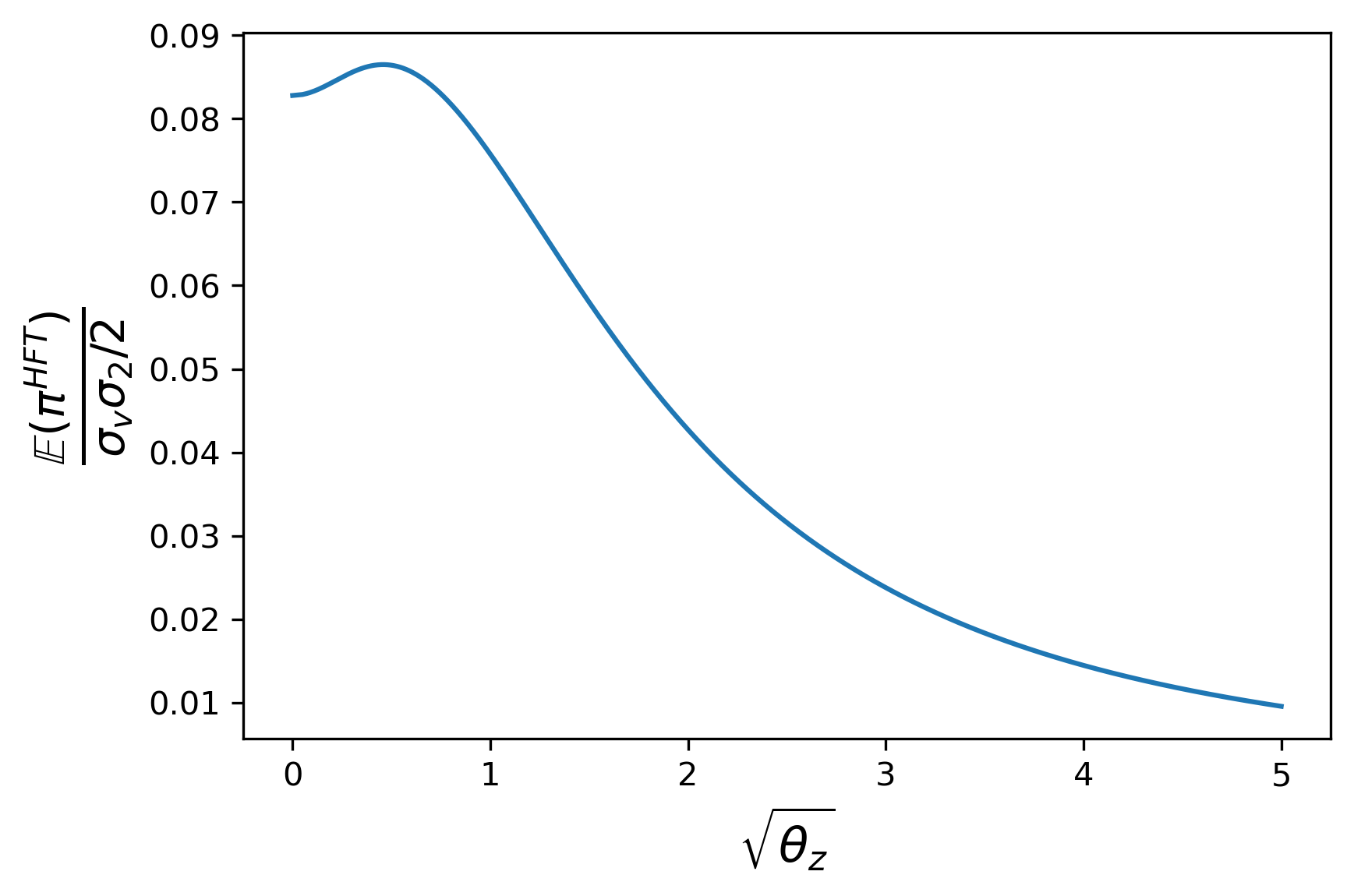}
    }
   \subcaptionbox{HFT's profit, $\theta_1=0.2$.}{
    \includegraphics[width = 0.27\textwidth]{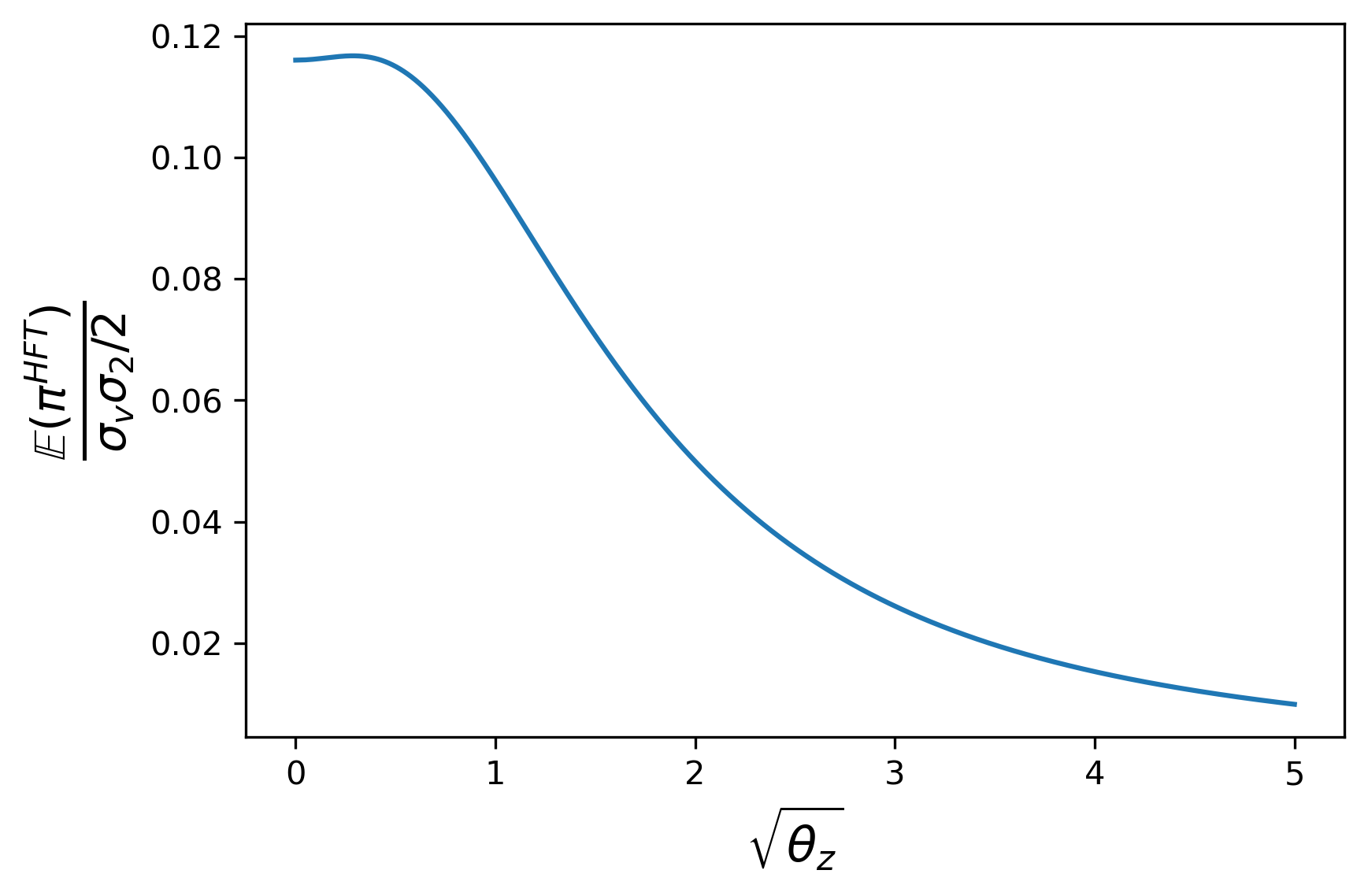}
    }
   \subcaptionbox{HFT's profit, $\theta_1=1$.}{
    \includegraphics[width = 0.27\textwidth]{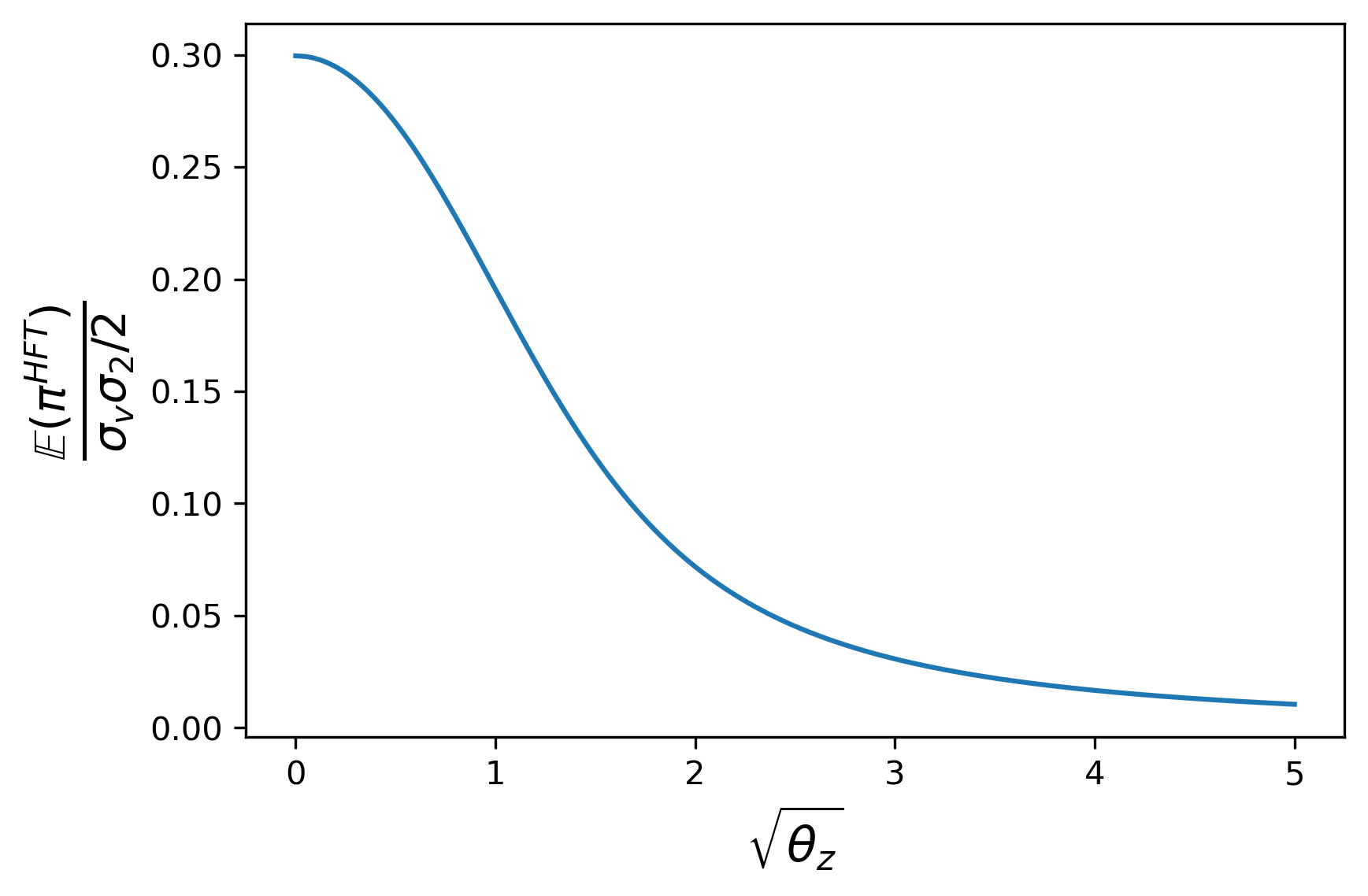}
    }
        \caption{How HFT's actions and profits change with $\theta_z$.}
    \label{figthetaz-HFT}
\end{figure}

\begin{figure}[!htbp]
    \centering
    \subcaptionbox{IT's action, $\theta_1=0.12\in(0,\frac{2\sqrt{3}-3}{3}].$}{
    \includegraphics[width = 0.27\textwidth]{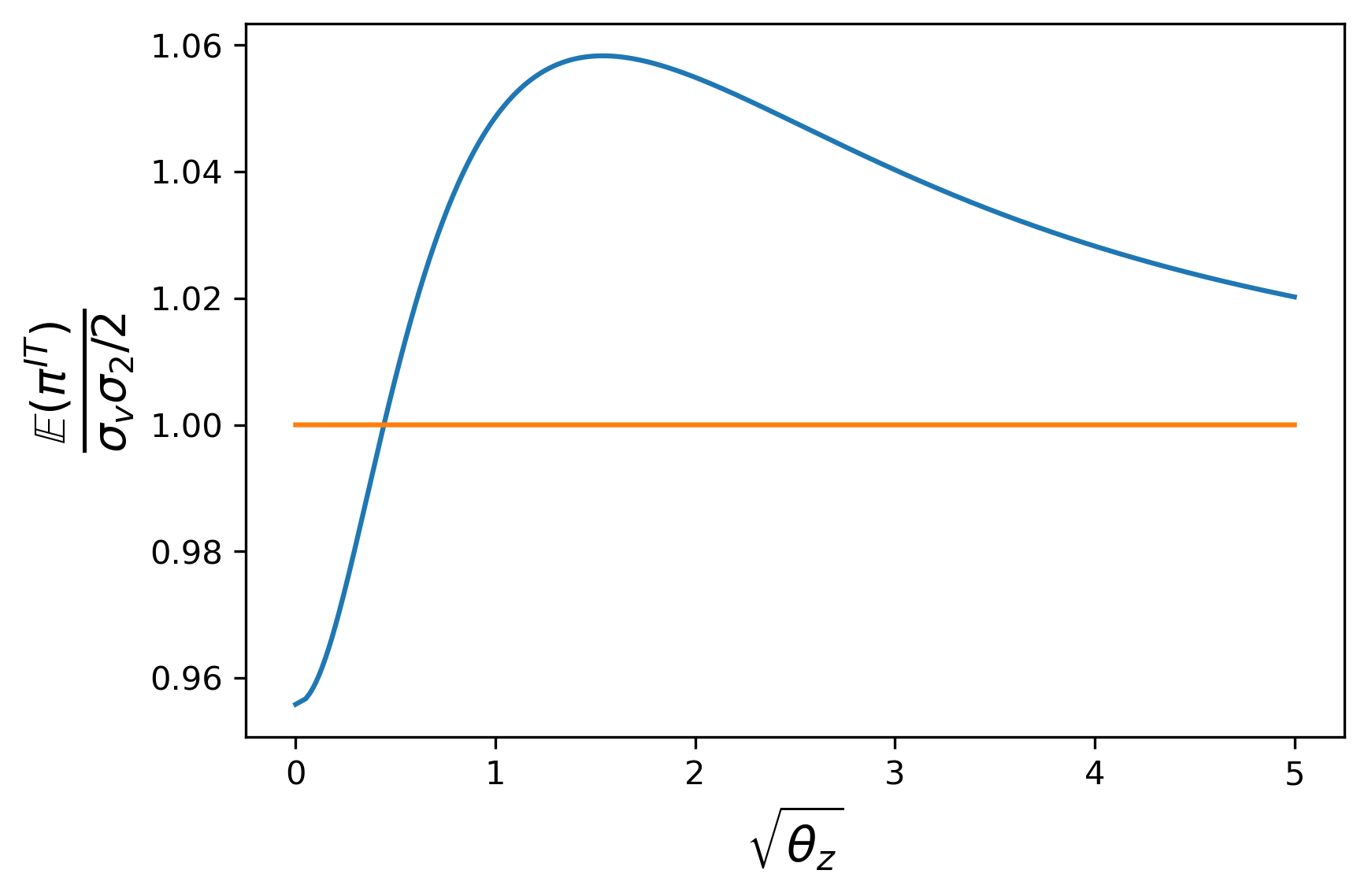}
    }
  \subcaptionbox{IT's action, $\theta_1=0.2\in(\frac{2\sqrt{3}-3}{3},\frac{1}{2})$.}{
    \includegraphics[width = 0.27\textwidth]{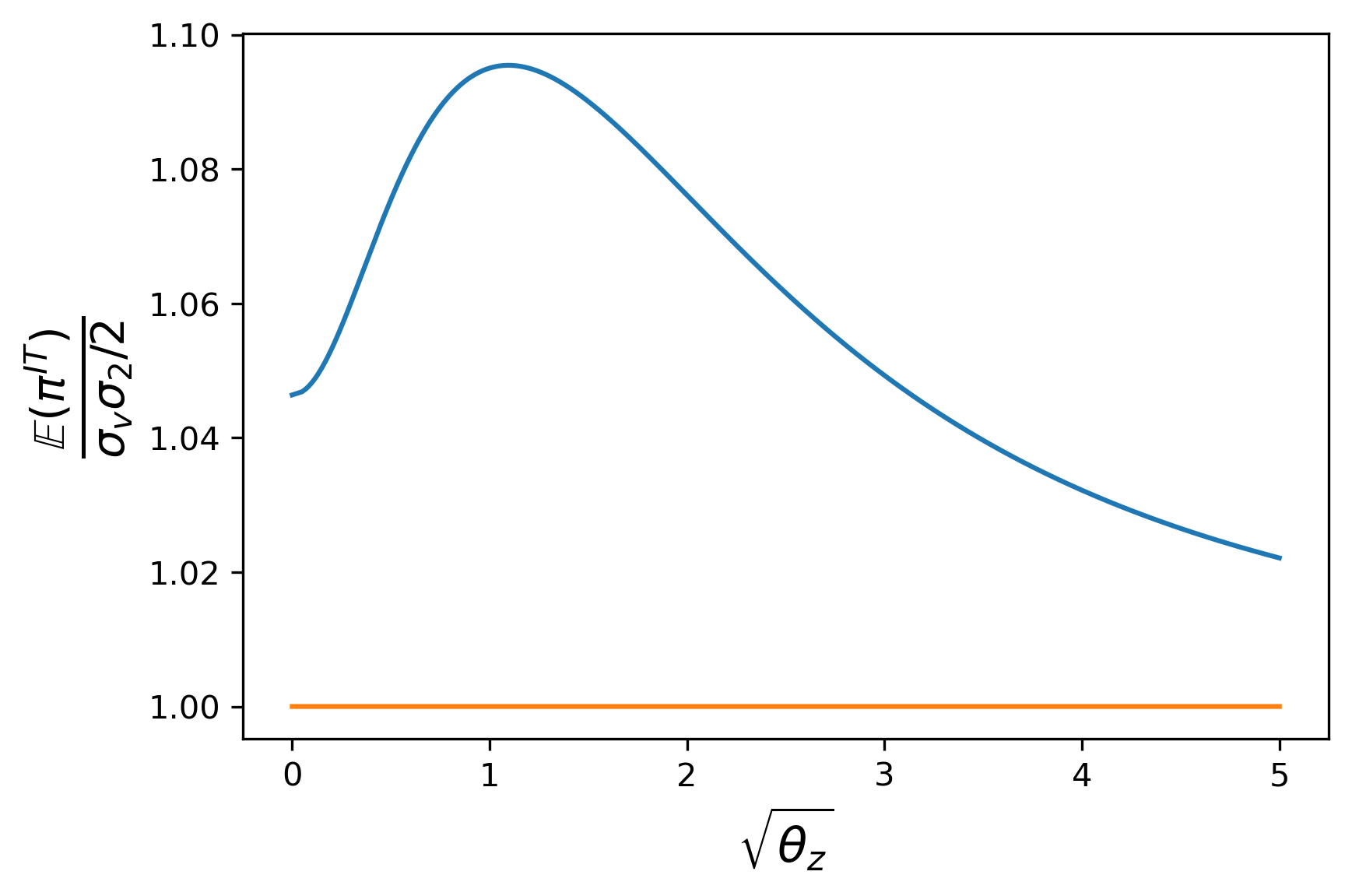}
    }
  \subcaptionbox{IT's action, $\theta_1=1\in[\frac{1}{2},+\infty)$.}{
    \includegraphics[width = 0.27\textwidth]{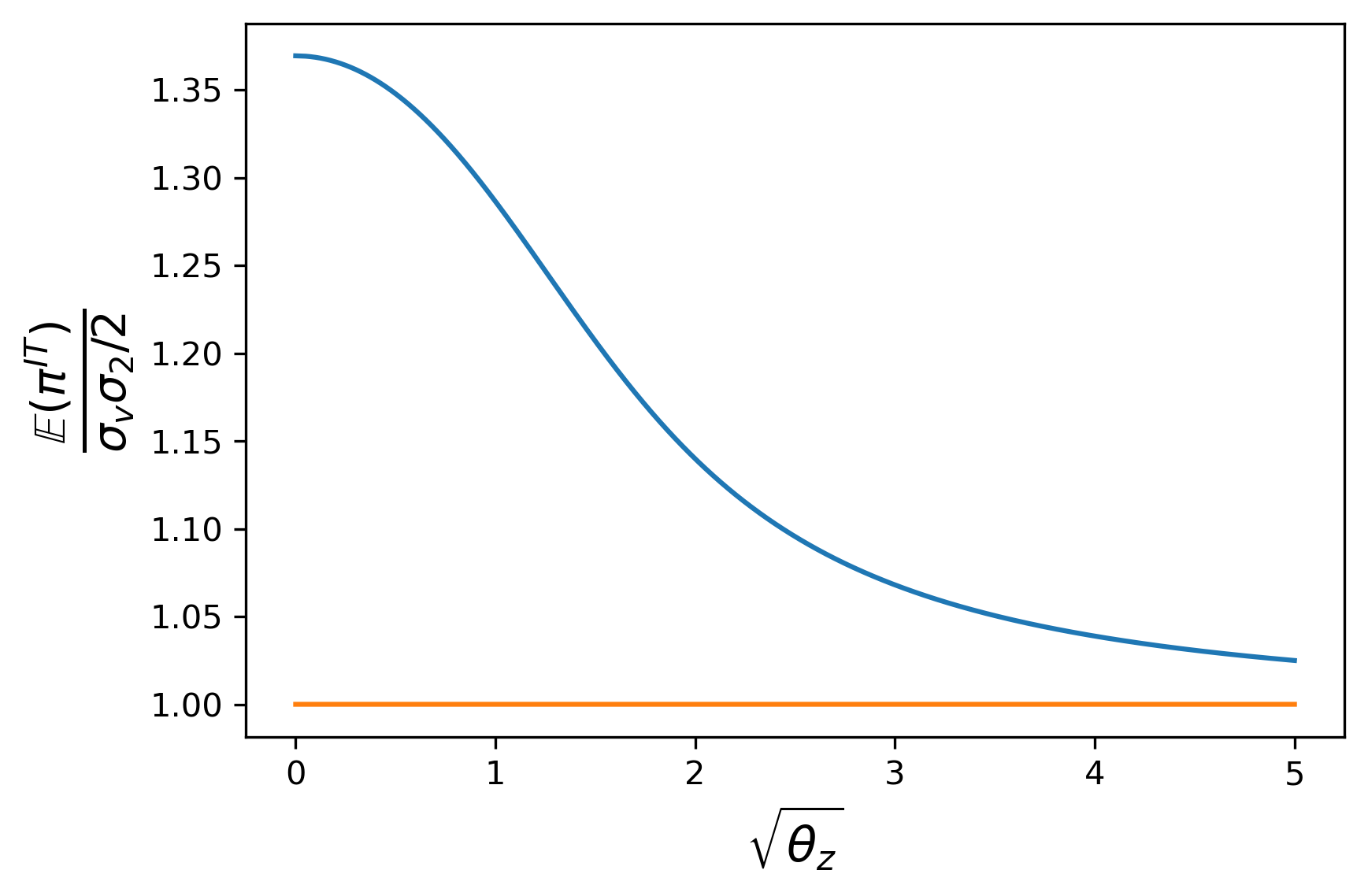}
    }
        \caption{How IT's actions (profits) change with $\theta_z$.}
    \label{figthetaz-IT}
\end{figure}

Next, we do comparative static analyses w.r.t. $\theta_z$: $\theta_z$ changes in $[0,25],$ given $\theta_1=0.12,0.2\text{ and }1.$ For HFT, the trading intensity $\beta^*$ decreases with $\theta_z$, as shown in (a)-(c) of Figure \ref{figthetaz-HFT}. Surprising results appear for HFT's expected profit. It is usually considered that HFT's profit should decrease if she receives a less accurate signal. However, as shown in (d) of Figure \ref{figthetaz-HFT}, it increases with $\theta_z$ when both $\theta_1$ and $\theta_z$ are relatively small.  It is because IT trades more as $\theta_z$ gets larger ((a) in Figure \ref{figthetaz-IT}), which brings more price impact for HFT to trade on.
\newpage

Figure \ref{figthetaz-IT} shows the trend of $\theta_z$'s impact on IT, corresponding to the results in Proposition \ref{ITaction}. 
As the signal becomes noisier, HFT's action $\beta^*$ decreases. On the one hand, HFT's time-1 trading causes less impact. On the other hand, her time-2 trading shares fewer transaction costs for IT. When the first effect exceeds the second one, IT's action and profit increase with $\theta_z$, otherwise, they decrease with it.

\section{When HFT Predicts the Aggregate Order}
In this section, we discuss an extension of the original model that HFT can predict $i+u_2,$ i.e., the signal received by HFT is
\begin{equation*}
\hat{y}=i+u_2+z,
\end{equation*}
where $z\sim N(0,\sigma_z^2),$ is the noise independent of other random variables. In this case, similar to Theorem \ref{mainthm}, when $\theta_1>0$, there exists a unique linear equilibrium where HFT front-runs. However, IT is always benefited by HFT.

When HFT predicts $i$, rather than $i+u_2$, the correlation of $y_1$ and $y_2$ is smaller. Since the time-2 price is 
$$\mathbb{E}(v|y_1,y_2)=\mu_1y_1+\mu_2y_2,$$ 
a smaller correlation brings a larger price impact coefficient $\mu_1,$ which makes it possible that HFT harms IT.

Li (2018) \cite{li2018high} models a market with normal-speed noise traders, i.e., $\theta_1=0$, where front-running HFTs predict $i+u_2$. The author concludes that IT is always harmed by HFT. The key points which make our conclusions different are: (1) \cite{li2018high} models stale dealers who quote linearly with unchanged impact coefficients, while we assume that dealers quote according to the weak-efficiency condition; (2) \cite{li2018high} assumes $\theta_1=0$.
Thus \cite{li2018high}'s results and our results can be seen as mutually complementary, explaining how front-running HFT affects IT when confronted with different liquidity providers and different sizes of market noise.

\section{Conclusion}
We study the influences of a front-running HFT on a large informed trader in various situations, where the size of high-speed noise trading and prediction accuracy differ. Since HFT takes liquidity away as well as provides liquidity back, she has a two-sided effect on the large trader. When the high-speed noise trading is sufficient, even if the large trader's intention is perfectly detected, she could be favored by HFT. Without enough noise shelter from the market, the large trader is benefited only when HFT's prediction is relatively vague. 

\section*{Acknowledgement}
Financial supports from National Natural Science Foundation of China under Grants No. 11971040 and The Fundamental Research Funds for the Central Universities, Peking University
are gratefully acknowledged.



\bibliographystyle{elsarticle-num}
\bibliography{procs-template-generic}









\end{document}